\documentclass{article}
\usepackage{spconf}
\usepackage{amsmath}
\usepackage{dsfont}
\usepackage{graphicx, graphics, theorem, times, amsfonts, amsmath, amssymb, cite}
\usepackage{tikz}
\usetikzlibrary{shapes,arrows}
\usepackage{pgfplots,comment}
\usepackage[x11names]{xcolor}
\usepackage{hyperref}
\usepackage{multirow}
\usepackage{rotating}
\usepackage{cleveref, subcaption}
\usepackage{bm}
\input{mysymbol.sty}
\usepackage[utf8]{inputenc}
\usepackage{enumitem}
\usepackage{algorithm,algpseudocode}
\usepackage{array}
\usepackage{moresize} 

\newcolumntype{P}[1]{>{\centering\arraybackslash}p{#1}}

\newcommand{\QED}{\hfill\ensuremath{\blacksquare}}

\newtheorem{myproposition}{\hspace{-1pt}\bf Proposition}

\def \bi {\begin{itemize}\item}

\def \ei {\end{itemize}}
\def \vec {\textnormal{vec}}
\def \tr {\textnormal{trace}}

\newtheorem{assumption}{\bf Assumption}


     \makeatother

     \makeatletter
     \renewcommand\small{%
     \@setfontsize\small\@ixpt{11}%
     \abovedisplayskip 3.5\p@ \@plus3\p@ \@minus2\p@
     \abovedisplayshortskip \z@ \@plus2\p@
     \belowdisplayshortskip 3\p@ \@plus2\p@ \@minus2\p@
     \def\@listi{\leftmargin\leftmargini
     \topsep 4\p@ \@plus2\p@ \@minus2\p@
     \parsep 2\p@ \@plus\p@
     \@minus\p@
     \itemsep
     \parsep}%
     \belowdisplayskip
     \abovedisplayskip
     }
     \makeatother

\title{Learning Product Graphs from Two-dimensional Stationary Signals
}

\name{Andrei Buciulea$^{\star}$ \qquad Bishwadeep Das$^{\dagger}$  \qquad Elvin Isufi$^{\dagger}$ \qquad Antonio G. Marques$^{\star}$\thanks{Part of this work was supported by the TU Delft AI Labs programme, the NWO OTP GraSPA project (No. 19497), the NWO VENI project (No. 222.032), the Spanish AEI (10.13039/501100011033) under grant PID2022-136887NBI00, and the Community of Madrid within the ELLIS Unit Madrid framework, the IDEA-CM (TEC-2024/COM-89), and the CAM-URJC F1180 (CP2301) grants. Corresponding author: antonio.garcia.marques@urjc.es.}}

\address{$^{\star}$ Dept. of Signal Theory and Communications, Rey Juan Carlos University, Madrid, Spain \\
         $^{\dagger}$ Delft University of Technology, Delft, The Netherlands}

\begin{document}

\ninept

\maketitle
\vspace{-0.1cm}
\begin{abstract}
\vspace{0.3cm}
Graph learning aims to infer a network structure directly from observed data, enabling the analysis of complex dependencies in irregular domains. Traditional methods focus on scalar signals at each node, ignoring dependencies along additional dimensions such as time, configurations of the observation device, or populations. In this work, we propose a graph signal processing framework for learning graphs from \emph{two-dimensional signals}, modeled as \emph{matrix graph signals} generated by joint filtering along both dimensions. This formulation leverages the concept of graph stationarity across the two dimensions and leverages product graph representations to capture structured dependencies. Based on this model, we design an optimization problem that can be solved efficiently and provably recovers the optimal underlying Kronecker/Cartesian/strong product graphs. Experiments on synthetic data demonstrate that our approach achieves higher estimation accuracy and reduced computational cost compared to existing methods.
\end{abstract}
\vspace{-0.1cm}
\begin{keywords}
Graph learning, network-topology inference, product graph, two-dimensional signals, graph-stationary signals
\end{keywords}

\vspace{-0.3cm}
\section{Introduction}\label{sec:introduction}
\vspace{-0.2cm}

Graphs are powerful for representing and analyzing signals on irregular domains. 
They model relationships between entities (nodes) and help extract insights from 
data (signals) observed over them. When the structure is unknown, graph learning 
(or network topology inference) aims to infer the graph directly from data. Early approaches were based on statistical methods~\cite[Ch.~7.3.1]{kolaczyk2009book}, including correlation- and partial-correlation-based techniques, which ultimately led to the development of the widely used graphical Lasso~\cite{meinshausen06,kolaczyk2009book}. Later, advances inspired by \emph{graph signal processing (GSP)} introduced models that explicitly incorporated the relationship between graph structure and observed signals.  Examples include smoothness-based models\cite{dong2016learning}, sparse structural equation models\cite{cai2013inference}, and stationarity-based approaches~\cite{egilmez2017graph,segarra2017network,buciulea2025polynomial}, which generalized classical models from regular domains (e.g., time series) to graphs. Building on these foundations, research has advanced toward more complex problems, considering the estimation of multiple graphs (time-varying graphs, task-specific, or online graph estimation)~\cite{kalofolias2017learning,danaher2014joint,navarro2022joint,sardellitti2021online,buciulea2025online}, handling perturbations in the graph and the signals (hidden nodes, missing data, or signal denoising)~\cite{chandrasekaran2012latent,chang2019graphical,buciulea2022learning,zhang2024joint}, and capturing higher-order interactions~\cite{sc_barbarosa,young2021hypergraph,buciulea2024learning}. 

Despite this progress, most graph learning frameworks produce graphs that capture dependencies along only a single dimension between the entities considered as nodes. Here, each node has a scalar observation, and multiple realizations of such signals are available. These realizations may correspond to time samples, repetitions of an experiment, or data collected from multiple individuals (e.g., brain activity from different patients) \cite{hallac2017network,Kalofolias2016inference_smoothAISTATS16,dong2019learning}. Depending on the type of signals and the scenario considered, one or multiple graphs can be estimated from these realizations. However, in many applications, a more appropriate representation is a \emph{two-dimensional signal}, where information is indexed both by the node (entities) and by the realization dimension (temporal, experimental, or population-level). Consequently, treating all realizations as a single dimension disregards the relationships across realizations, thereby limiting the interpretability and comprehensiveness of the learned graphs in contexts where dependencies across multiple dimensions are significant.

To model the dependencies across multiple dimensions simultaneously, recent research considers each observation to be indexed not only by a node (e.g., a movie) but also by an additional dimension such as time instants, experiments or users, resulting in two-dimensional or multiway graph signals. Thus, a (movie-user) node in a recommender system establishes the relationship between a user and a movie jointly as an entity. Several approaches have been developed to capture these structured dependencies. The \emph{bigraphical Lasso} proposes an estimator for precision matrices based on the Cartesian product of graphs to jointly capture correlations across rows and columns~\cite{kalaitzis2013bigraphical}, while \emph{TERRALasso} generalizes this idea to multi-way tensor data with a sparse Kronecker-sum structure~\cite{greenewald2019tensor}. \emph{EiGLasso} further improves scalability by exploiting eigendecomposition and Newton-based approximations for estimating sparse Kronecker-sum precision matrices~\cite{greenewald2019eiglasso}, and \emph{DNNLasso} ensures interpretability and computational efficiency via a diagonally non-negative graphical Lasso~\cite{lin2024dnnlasso}. Other works leverage smoothness priors and structured graph products—including Cartesian, Kronecker, and strong products—to learn graphs along multiple dimensions for capturing non-separable dependencies~\cite{lodhi2020learning,shi2025learning,kadambari2020learning,shi2024learning,kadambari2021product,zhang2023product}.

\noindent\textbf{Our contribution.} In this work, we adopt a GSP perspective to learn graphs from two-dimensional signals, assuming signal stationary with respect to the underlying network. Our contribution is to model stationary signals as the output of a joint filter that encodes the underlying graph structures along each dimension while acting on white Gaussian noise, thus, extending stationarity to two-dimensional graph signals and exploiting the Kronecker structure of the underlying graph. Based on this model, we develop an algorithm that leverages the Kronecker structure to learn graphs via a separable optimization problem which links to existing formulations for graph recovery from multidimensional signals. Finally, we validate our approach through extensive synthetic experiments against state-of-the-art baselines, demonstrating improvements in both estimation accuracy and computational efficiency.

\section{Preliminaries: Learning graphs from one-dimensional stationary signals}\label{S:ProblemSetup}

A \emph{graph signal} is a magnitude (feature) that varies over the nodes of a graph. In the graph-learning context, the goal is to infer the underlying graph structure (i.e., the pairwise relations among the nodes) from observed graph signals. Mathematically, this entails estimating $\bbS \in \mathbb{R}^{N \times N}$, the so-called \emph{graph shift operator} (GSO) matrix, from $R$ signals in $\bbX = [\bbx_1,\ldots,\bbx_R] \in \mathbb{R}^{N \times R}$, with $N$ denoting the number of nodes in the graph. The GSO encodes the graph topology and is commonly chosen as the adjacency matrix, the graph Laplacian, or their generalizations~\cite{djuric2018cooperative}. 

Stationary graph signals provide a model for linking data to graph structure. A stationary graph signal can be expressed as $\bbx = \bbH \bbw$, where $\bbw$ is standard normally distributed vector, and $\bbH$ is a \emph{graph filter} of the form $\bbH = \sum_{l=0}^{L-1} h_l \bbS^l$,
i.e., a polynomial in the GSO with coefficients $\{h_l\}$. The covariance matrix of a stationary signal, $\bbC_{\bbx} = \mathbb{E}[\bbx \bbx^\top]$, is also a polynomial in $\bbS$~\cite{marques2017stationary}. Consequently, $\bbC_{\bbx}$ and $\bbS$ share the same eigenvectors and commute~\cite{segarra2017network,shafipour2020online,buciulea2022learning}, i.e.,
$\bbC_{\bbx}\bbS = \bbS\bbC_{\bbx}$. This motivates graph learning problem based on stationarity, where given the true covariance $\bbC_{\bbx}$, one seeks to recover a sparse GSO by solving
\begin{alignat}{2}\label{E:org_prob0}
    \min_{\bbS \in \ccalS} \quad & \|\bbS\|_1 
     \;\;\;\text{s. t.} \quad & \bbC_{\bbx} \bbS = \bbS \bbC_{\bbx}, 
\end{alignat}
where the $\ell_1$-norm promotes sparsity in the estimated graph, which is a typical assumption for real-world networks, and the commutativity constraint ensures that the signal is stationary on the estimated graph. $\ccalS$ denotes the set of valid GSOs, e.g., adjacency matrices: $\ccalS = \big\{\bbS \in \mathbb{R}^{N \times N} \,\big|\, 
\bbS = \bbS^\top, S_{ij} \ge 0, S_{ii} = 0, \bbS \mathbf{1} = \mathbf{1} \big\}$,
where the matrices are symmetric (undirected graphs), nonnegative, and have zero diagonal (no self-loops), and the last constraint avoids the trivial zero solution.

\section{Learning graphs from Two-Dimensional Stationary Signals}

\emph{Two-dimensional} graph signals naturally arise in applications such as spatio-temporal processes, images, or multi-modal data~\cite{yan2024signal,stanley2020multiway}. Such signals can be represented as a matrix $\bbY \in \mathbb{R}^{P \times Q}$, where the two dimensions capture distinct aspects of the data. For example, in spatio-temporal signals, $P$ may correspond to spatial nodes and $Q$ to time instances; more generally, the dimensions can represent any pair of modalities. When multiple realizations are available, the data can be organized as a tensor $\underline{\bbY} \in \mathbb{R}^{P \times Q \times R}$, with $\bbY_r\in \mathbb{R}^{P \times Q}$ denoting the $r$-th slab of the tensor. 

\vspace{0.1cm}
\noindent\textbf{General vectorized model.}  
If each $(p,q)$ pair is considered as a node, the most general approach is to vectorize $\bbY$ into $\bby = \mathrm{vec}(\bbY) \in \mathbb{R}^{PQ}$, consider a one-dimensional stationary graph signal model, and estimate the GSO $\bbS_{PQ} \in \mathbb{R}^{PQ \times PQ}$ without imposing any additional structure on the graph. Specifically, modelling the observations as stationary implies that $\bby = \bbH_{PQ}\bbw$ with $\bbw \sim \mathcal{N}(\mathbf{0},\bbI_{PQ})$,
where $\bbH_{PQ}$ is a polynomial in the shift operator $\bbS_{PQ} \in \mathbb{R}^{PQ \times PQ}$ and $\bbI_{PQ}\in \reals^{PQ \times PQ}$ denotes the identity matrix. In this formulation, $\bbS_{PQ}$ captures all possible relationships across both dimensions, treating each pair $(p,q)$ as a node in the graph. The covariance $\bbC_\bby = \mathbb{E}[\bby\bby^\top]$ commutes with $\bbS_{PQ}$, allowing the graph topology to be inferred by solving Problem~\eqref{E:org_prob0}. However, this approach suffers from poor scalability. First, the number of unknowns grows quadratically with $PQ$, which can be prohibitive even for estimation algorithms that run in polynomial time. Second, accurately estimating $\bbC_{PQ}$ is difficult, since the sample covariance $\frac{1}{R}\sum_{r=1}^R \mathrm{vec}(\bbY_r)\mathrm{vec}(\bbY_r)^T$ is typically low rank, requiring either a very large $R$ or strong regularization.

\vspace{0.1cm}
\noindent\textbf{Structured product-graph model.}  
To address these limitations, we consider a structured approach by assuming that the observed signals are stationary on a product graph. Specifically, let $G_P = (\mathcal{V}_P, S_P)$ and $G_Q = (\mathcal{V}_Q, S_Q)$ be two graphs with $|\mathcal{V}_P| = P$ and $|\mathcal{V}_Q| = Q$ nodes, respectively, that capture relationships across dimensions $P$ and $Q$. The product graph of $G_P$ and $G_Q$, denoted by $\diamondsuit$, is defined as $G_{PQ} = G_P \diamondsuit G_Q = (\mathcal{V}, A_\diamondsuit)$, where $\diamondsuit$ represents either the Kronecker, Cartesian, or strong product \cite{sandryhaila2014big}. A key property of all these constructions is that the eigenvectors of the product GSO are given by $\bbV_Q\otimes\bbV_P$, the Kronecker product of the eigenvectors of $\bbS_P$ and $\bbS_Q$ \cite{sandryhaila2014big}.  
In our setting, this is relevant because the assumption of stationarity on the product graph implies that the covariance matrix $\bbC_\bby$ shares its eigenvectors with the Kronecker product GSO $\bbS_{PQ} = \bbS_P \otimes \bbS_Q$. This observation motivates the following topology identification problem:
\begin{alignat}{2}\label{E:org_prob1}
    \min_{\bbS_P \in \ccalS, \, \bbS_Q \in \ccalS} \quad & \|\bbS_P\|_1 + \beta\|\bbS_Q\|_1 \\
    \text{s. t.} \quad & \bbC_\bby (\bbS_Q \otimes \bbS_P) = (\bbS_Q \otimes \bbS_P)\bbC_\bby, \nonumber
\end{alignat}
where the objective promotes sparse solutions by minimizing the sum of the $\ell_1$ norms of $\bbS_P$ and $\bbS_Q$ with $\beta>0$. 
Sparsity in both $\bbS_P$ and $\bbS_Q$ directly implies that the GSO of the product graph will also be sparse. The commutativity constraint between $\bbC_\bby$ and $\bbS_P \otimes \bbS_Q$ follows from the assumption of stationarity on the product graph, since both matrices are simultaneously diagonalizable. Incorporating this additional structure reduces the number of optimization variables from $P^2Q^2$ (in the fully general case) to $P^2 + Q^2$, thereby improving tractability. Nonetheless, the problem remains nonconvex due to the Kronecker product in the commutativity constraint. Alternating optimization schemes are typically employed to address this and often yield good empirical results. However, recovery guarantees are difficult to establish, the large covariance matrix must still be estimated, and computations continue to involve handling large $PQ \times PQ$ matrices.

\vspace{0.1cm}
\noindent\textbf{Dimension-wise covariance formulation.}  
In this section, we adopt a more structured signal model that will lead to a more tractable optimization, bypassing some of the issues present when solving \eqref{E:org_prob1}. The critical point is to assume that the two-dimensional signal is generated by a particular graph filtering process, as described next. 
\vspace{-0.2cm}
\begin{assumption}\label{E: assumption1}
Given GSOs $\bbS_P \in  \reals^{P \times P}$ and $\bbS_Q\in \reals^{Q \times Q}$, and polynomial graph filters $\bbH_P(\bbS_P)$ and $\bbH_Q(\bbS_Q)$, the two-dimensional graph signal $\bbY \in \reals^{P \times Q}$ is generated as
\begin{align}\label{E:sig_gen}
    \bbY = \bbH_P(\bbS_P) \, \bbW \, \bbH_Q(\bbS_Q),
\end{align}
where the entries of the noise matrix $\bbW \in \reals^{P \times Q}$ are i.i.d. 
\end{assumption}
Here, $\bbH_P(\bbS_P)$ and $\bbH_Q(\bbS_Q)$ are polynomial filters in their respective GSOs, and for simplicity we refer to them as $\bbH_P$ and $\bbH_Q$.
Assumption \ref{E: assumption1} states that the observations are generated through distinct filtering processes along each dimension. In a spatiotemporal setting, for instance, the dependencies across time may differ significantly from those across space, requiring different diffusions. Similarly, in recommender systems \cite{huang2018rating}, the user and item dimensions may require different filters due to the heterogeneous nature of their underlying graphs. Moreover, the data generation model in \eqref{E:sig_gen} parallels the stationary time-vertex signal processing framework in \cite{loukas2019stationary}, even though one of the two dimensions need not be temporal. If we interpret the Laplacian of the temporal graph in \cite{loukas2019stationary} as the shift operator in one dimension, the vectorized signals $\text{vec}(\bbY)$ satisfy the criterion of joint wide-sense stationarity \cite[Def. 1]{loukas2019stationary}. The matrix $\bbY$ also satisfies the criterion of a jointly wide-sense stationary process, provided that both graphs are connected.\footnote{The proofs follow directly from \cite{loukas2019stationary}.}

Signals satisfying Assumption \ref{E: assumption1} clearly give rise to covariances diagonalized by $\bbV_Q \otimes \bbV_P$. To see this, note that vectorizing $\bbY$ in \eqref{E:sig_gen} yields
\begin{align}\label{E:sig_gen_vec}
\bby = \big(\bbH_Q \otimes \bbH_P\big)\vec(\bbW).
\end{align}
As a result, we have
\begin{align}\label{E:cov_H_kron}
\bbC_\bby = \mathbb{E}[\bby \bby^\top]
= \bbH_Q\bbH_Q^\top \otimes \bbH_P\bbH_P^\top
= \bbH_Q^2 \otimes \bbH_P^2.
\end{align}
Since $\bbH_Q^2$ is a polynomial in $\bbS_Q$, its eigenvectors are $\bbV_Q$. Similarly, the eigenvectors of $\bbH_P^2$ are $\bbV_P$. Hence, the eigenvectors of $\bbC_\bby$ in \eqref{E:cov_H_kron} are $\bbV_Q \otimes \bbV_P$. This implies that when signals follow the model in Assumption \ref{E: assumption1}, the GSOs $\bbS_P$ and $\bbS_Q$ can be identified by solving \eqref{E:org_prob1}.



Interestingly, the structure of \eqref{E:sig_gen} enables the design of a more efficient approach for estimating the GSO. To this end, we first present the following result as a proposition.
\begin{myproposition}\label{Proposition}
Let $\bbY$ be a two-dimensional graph signal satisfying Assumption~\ref{E: assumption1} and define matrices $\bbC_P=\mathbb{E}[\bbY \bbY^\top]$ and $\bbC_Q=\mathbb{E}[\bbY^\top\bbY]$. Then, it holds that:
\begin{itemize}
\item[i)] $\bbC_P\!=\operatorname{tr}\!\big(\bbH_Q^2\big)\! \bbH_P^2$, which is a polynomial in $\bbS_P$.
\item[ii)] $\bbC_Q\!=\operatorname{tr}\!\big(\bbH_P^2\big)\! \bbH_Q^2$, which is a polynomial in $\bbS_Q$.
\end{itemize}
\end{myproposition}
\noindent\textbf{Proof.} See Appendix. \QED

The result in the proposition allows us to compute (estimate) the covariance matrices separately and formulate the graph-learning problem as 
\begin{align}\label{P:separable_sum}
\begin{split}
&\underset{\bbS_P\in\ccalS,\,\bbS_Q\in\ccalS}{\text{min}} \quad \|\bbS_P\|_1 + \beta\|\bbS_Q\|_1\\
\text{s.t.} \quad & \bbC_P \bbS_P = \bbS_P \bbC_P, \quad 
\bbC_Q \bbS_Q = \bbS_Q \bbC_Q.
\end{split}
\end{align}
where the fact that $\bbC_P$ and $\bbC_Q$ are polynomials in $\bbS_P$ and $\bbS_Q$ is encoded in the (separate) commutativity constraints. This formulation is convex in $\{\bbS_P, \bbS_Q\}$ and separable, allowing each graph to be optimized independently. In addition, the number of constraints is significantly smaller than in \eqref{E:org_prob1}, which reduces computational complexity. In practice, the covariances are estimated from $R$ observations as $\hbC_P = \sum_{r=1}^R \bbY_r \bbY_r^\top$ and $\hbC_Q = \sum_{r=1}^R \bbY_r^\top \bbY_r$. Since these covariance matrices are now much smaller, modest values of $R$ are sufficient to obtain reliable estimates. The formulation in \eqref{P:separable_sum} is a special case of eq (45) in \cite{einizade2023learning} for $n=2$, $\alpha_1=0.5$, and $\alpha_2=2\beta$.

Having established the different problem formulations, we now summarize them along with their corresponding solution strategies. Problem~\eqref{E:org_prob0} is convex in $\bbS$ and can be solved using standard off-the-shelf solvers or more efficient methods such as ADMM~\cite{BoydADMM}, resulting in an unstructured learned graph. The main drawback in this case is the high dimensionality, which complicates both the computational complexity and the accurate estimation of covariance matrices. In contrast, Problem~\eqref{E:org_prob1} imposes a product graph structure. This reduces the number of optimization unknowns, but introduces a non-convex (Kronecker product) constraint that must be handled via alternating optimization~\cite{gorski2007biconvex}. While this approach lowers computational complexity, the estimation of the high-dimensional covariance matrix remains an issue. Finally, by leveraging additional structure in the graph signals, Problem~\eqref{P:separable_sum} yields a convex, low-dimensional optimization that can be solved separately for each dimension, thereby bypassing the need to estimate the full covariance matrix.

\section{Numerical experiments}\label{S: Numerical Experiments}
\begin{figure*}[t]
    \centering
    \begin{subfigure}[b]{0.32\textwidth}
        \includegraphics[width=\textwidth]{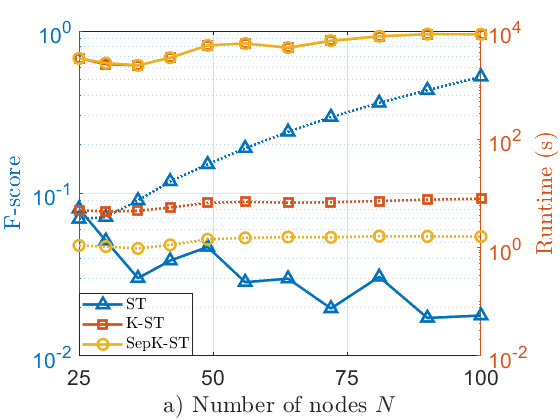}
    \end{subfigure}
    \begin{subfigure}[b]{0.32\textwidth}
        \includegraphics[width=\textwidth]{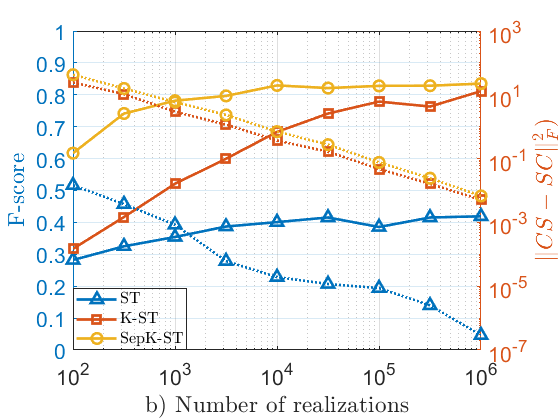}
    \end{subfigure}
    \begin{subfigure}[b]{0.32\textwidth}
        \includegraphics[width=\textwidth]{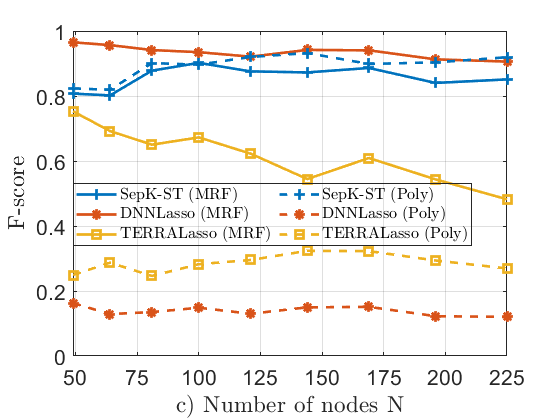}
    \end{subfigure}
    \caption{Performance comparison of three proposed approach variants. a) F-score (solid lines, left y-axis) and runtime in seconds (dotted lines, right y-axis) plotted against the total number of nodes. b) F-score (solid lines, left y-axis) and commutativity values of the estimated graphs (dotted lines, right y-axis) as a function of the number of samples.  c) F-score comparison of the proposed approach (SepK-ST) against two methods from the literature, as the graph size increases, under two different signal generation models: $\bbC_{MRF}$ and $\bbC_{Poly}$.}
    \label{fig:results}
\end{figure*}
We evaluate the performance of the proposed formulations on synthetic data and compare them against existing approaches from the literature. The methods considered are: spectral templates (ST), which solves the problem in \eqref{E:org_prob0}; Kronecker-ST (K-ST), which addresses the problem in \eqref{E:org_prob1}; separate Kronecker-ST (SepK-ST), which solves \eqref{P:separable_sum}; DNNLasso, the method proposed in \cite{lin2024dnnlasso}; and TERRALasso, the method introduced in \cite{greenewald2019tensor}.
For evaluation, we focus on unweighted graphs and employ the F-score, defined as 
\begin{equation} 
\textnormal{F\text{-}score} = 2 \cdot \frac{\textnormal{precision} \cdot \textnormal{recall}}{\textnormal{precision} + \textnormal{recall}}, 
\end{equation}
where $precision$ denotes the percentage of estimated edges that are edges of the ground-truth graph and $recall$ the percentage of existing edges that were correctly estimated. In all experiments, both $\bbS_P$ and $\bbS_Q$ are generated as Erdős-Rényi random graphs with a node connection probability of $0.3$. Simulations are repeated over 100 independent graph realizations, and the mean F-score is reported. Implementations of the proposed approaches are carried out in \texttt{CVX}. More specific details about the implementation setup of the experiments can be found in the available code on GitHub\footnote{https://github.com/andreibuciulea/twodimST}.

\vspace{0.1cm}
\noindent \textbf{Accuracy vs. time complexity.} 
We compare ST, K-ST, and SepK-ST in terms of graph recovery accuracy and computational runtime. Fig.~\ref{fig:results}a) reports the F-score and runtime as the graph size increases, under the assumption of perfect covariance knowledge.
From the F-score results (solid lines), we observe that both K-ST and SepK-ST achieve identical recovery performance and clearly outperform ST, which fails to correctly capture the underlying structure and exhibits significantly lower F-scores. Both K-ST and SepK-ST explicitly exploit the two-dimensional structure of the signals, enabling accurate recovery of the Kronecker-structured graphs, whereas ST ignores this structure. Regarding runtime (dotted lines), all methods become slower as the graph increases, which is consistent with the higher complexity of larger problems. However, ST scales much worse than K-ST and SepK-ST as it requires estimating $(PQ)^2$ variables, whereas K-ST and SepK-ST only involve $P^2 + Q^2$ variables. Between K-ST and SepK-ST, the difference lies in implementation: although they optimize over the same number of variables, K-ST requires large matrix operations because of Kronecker products, which increases runtime and makes it slower than SepK-ST.  
This highlights the importance of leveraging the inherent structure of two-dimensional signals. SepK-ST provides a good balance between recovery performance and runtime.  

\vspace{0.1cm}
\noindent \textbf{Accuracy vs. commutativity.} 
We now evaluate the behavior of ST, K-ST, and SepK-ST when the covariance matrix is estimated from a limited number $R$ of signals. Fig.~\ref{fig:results}b) reports the F-score and commutativity of the estimated graphs as $R$ increases, fixing the graph size to $N=64$ nodes with $P=Q=8$. The F-scores (solid lines) show that SepK-ST outperforms ST and K-ST, achieving better graph recovery across different $R$. This is partly because smaller graphs require fewer samples for accurate estimation, while ST must handle a much larger graph and thus needs more samples. K-ST relies on the covariance of the vectorized two-dimensional signals, which also requires more samples to reach performance comparable to SepK-ST.  
We also assess commutativity, defined as $\|\bbC\hat{\bbS}-\hat{\bbS}\bbC\|_F^2$, where $\hat{\bbS}$ is the estimated Kronecker graph, i.e., how well the estimated graphs satisfy stationarity (lower implies better). ST consistently yields lower values than K-ST and SepK-ST, indicating that it identifies graphs on which the signals are stationary, but these graphs fail to capture the true two-dimensional structure. In contrast, SepK-ST shows decreasing commutativity values alongside increasing F-score as $R$ grows, demonstrating its ability to enforce stationarity while recovering model-consistent graphs.  
Lastly, K-ST attains lower commutativity values than SepK-ST, but the recovered graphs deviate more from the ground truth, suggesting that K-ST requires more samples to match SepK-ST. This discrepancy shrinks as $R$ increases, reducing the F-score performance gap. Overall, the results indicate that SepK-ST is the most reliable approach in sample-limited regimes, balancing stationarity and accurate recovery.

\vspace{0.1cm}
\noindent \textbf{Comparing baselines.} 
Here, we compare the proposed approaches against two state-of-the-art algorithms from the literature that also account for two-dimensional graph signals with product graph structure. To ensure fairness, we generate signals using two distinct models whose true covariances are given by $\bbC_{MRF} = (\alpha\bbI + \bbS)^{-1}$ and $\bbC_{Poly} = \big(\sum_{l=1}^{L} h_l \bbS^l \big)^2$, with $\alpha > 0$ sufficiently large to guarantee positive definiteness and $L = 3$. The number of samples is fixed to $10^6$.  
The F-score results (solid lines in Fig.~\ref{fig:results}c) show that DNNLasso achieves the best performance, closely followed by SepK-ST. TERRALasso is also able to recover the graph, but its accuracy decreases significantly as the graph size grows. This happens because $\bbC_{MRF}$ aligns perfectly with the modeling assumptions of DNNLasso, explaining its superior recovery performance.  
When signals are generated obeying $\bbC_{Poly}$, our method is the only one capable of consistently recovering the underlying graph while maintaining high F-score as the number of nodes increases. In contrast, DNNLasso and TERRALasso fail to capture the structure under this scenario.  
This suggests that our approach is not only accurate and computationally efficient, but it is also verstaile as it generalizes to graphs arising from different two-dimensional data models.




\section{Conclusions}\label{S: Conclusion}

In this work, we introduced a novel framework for learning graphs from two-dimensional stationary signals under a GSP perspective. After discussing a generic stationarity-based formulation, we developed a lower-complexity approach that models the overall graph as the Kronecker/Cartesian/strong product of two separate graphs. Finally, by imposing additional structure and modeling the observed signals as the output of a separate graph filtering across dimensions, we established explicit connections between dimension-specific covariances and their underlying graph shift operators. This allowed us to reformulate the topology identification task as two separable, tractable convex problems, each defined over a single dimension, while preserving the Kronecker structure of the model. 
Numerical experiments further demonstrated that the proposed approach achieves high recovery accuracy, scales more efficiently with graph size, and outperforms existing baselines in both accuracy and runtime.

\section{Appendix}\label{S: Appendix}

\noindent\textbf{Proof of Proposition \eqref{Proposition}.}  
We begin by vectorizing \eqref{E: assumption1}. This yields
\[
\bby = (\bbH_Q \otimes \bbH_P)\,\vec(\bbW),
\]
so that the theoretical covariance of $\bby$ is
\begin{equation}
\bbC_\bby = \mathbb{E}[\bby\bby^\top] = \bbH_Q^2 \otimes \bbH_P^2. 
\end{equation}

\medskip
\noindent\textbf{Matrix-form analysis.}  
Alternatively, let us work directly with the matrix $\bbY$. The covariance along dimension $P$ is
\[
\bbC_P = \mathbb{E}[\bbY\bbY^{\top}] 
= \mathbb{E}[\bbH_P \bbW \bbH_Q \bbH_Q^{\top} \bbW^{\top} \bbH_P^{\top}].
\]
Since 
\[
\mathbb{E}[\bbW \bbH_Q \bbH_Q^{\top} \bbW^{\top}] = \tr(\bbH_Q^2)\,\bbI,
\]
it follows that
\begin{equation}\label{covariances}
\bbC_P = \tr(\bbH_Q^2)\,\bbH_P^2, 
\qquad
\bbC_Q = \tr(\bbH_P^2)\,\bbH_Q^2.
\end{equation}

\medskip
\noindent\textbf{Relation between covariances.}  
Taking the Kronecker product of the two covariance matrices, we obtain
\begin{align}\label{Covariance relation}
\begin{split}
 \bbC_Q \otimes \bbC_P 
 &= \tr(\bbH_Q^2)\,\tr(\bbH_P^2)\,(\bbH_Q^2 \otimes \bbH_P^2) \\
 &= \|\bbH_Q\|_F^2 \,\|\bbH_P\|_F^2 \,\bbC_\bby,
\end{split}
\end{align}
which establishes the connection between $ \bbC_Q \otimes \bbC_P$ and $\bbC_\bby$.

\medskip
\noindent\textbf{Spectral characterization.}  
Let the eigendecompositions of the shift operators be
\[
\bbS_Q = \bbV_Q \bbLambda_Q \bbV_Q^{\top},
\qquad
\bbS_P = \bbV_P \bbLambda_P \bbV_P^{\top}.
\]
It follows that
\[
\bbH_Q^2 = \bbV_Q \bbH(\bbLambda_Q)^2 \bbV_Q^{\top},
\qquad
\bbH_P^2 = \bbV_P \bbH(\bbLambda_P)^2 \bbV_P^{\top},
\]
where $\bbH(\bbLambda_Q)$ and $\bbH(\bbLambda_P)$ are diagonal matrices containing the frequency responses.  

Substituting these into \eqref{covariances}, we see that $\bbC_Q$ and $\bbC_P$ share eigenvectors with $\bbS_Q$ and $\bbS_P$, respectively. Likewise, substituting into \eqref{Covariance relation} gives
\begin{align}
\bbC_Q \otimes \bbC_P 
= (\bbV_Q \otimes \bbV_P)\,
\big(\bbH(\bbLambda_Q)^2 \otimes \bbH(\bbLambda_P)^2\big)\,
(\bbV_Q^{\top} \otimes \bbV_P^{\top}).
\end{align}

Therefore, $\bbC_Q \otimes \bbC_P$ commutes with $\bbS_Q \otimes \bbS_P$. \QED

\vfill\pagebreak
\bibliographystyle{IEEEtran}
\bibliography{citations}

\end{document}